\documentclass[aps,pre,reprint,groupedaddress]{revtex4-2}

\usepackage{color}
\usepackage{amsmath}
\usepackage{graphicx}
\usepackage{booktabs}
\usepackage{url}

\begin{document}


\title{Palette diagram: A Python package for visualization of collective categorical data}


\author{Chihiro Noguchi}
\affiliation{Department of Mathematical and Computing Science, 
Tokyo Institute of Technology, 2-12-1 Ookayama, Meguro-ku, Tokyo, Japan}

\author{Tatsuro Kawamoto}
\affiliation{  Artificial Intelligence Research Center, 
  National Institute of Advanced Industrial Science and Technology, 
  2-3-26 Aomi, Koto-ku, Tokyo, Japan }


\date{\today}

\begin{abstract}
Categorical data, wherein a numerical quantity is assigned to each category (nominal variable), are ubiquitous in data science. 
A palette diagram is a visualization tool for a large number of categorical datasets, each comprising several categories. \\

\hfil \textbf{Github repository: https://github.com/palette-diagram/palette-diagram} \hfil
\end{abstract}


\maketitle

When dealing with only one categorical dataset, we only need a bar plot or pie chart. 
Even when we deal with a large set of categorical data, if the data have an intrinsic order, such as time series, we can use a stack plot or stream plot \cite{byron2008streamgraph}. 
A palette diagram is a visualization tool for a set of categorical data without an intrinsic order. 
Examples include a large set of count data (a contingency table with a large number of columns or rows) and a result of Bayesian inference in which we have a posterior probability distribution for each element of the input dataset. 

\begin{figure}[ht!]
  \centering
  \includegraphics[width= \columnwidth]{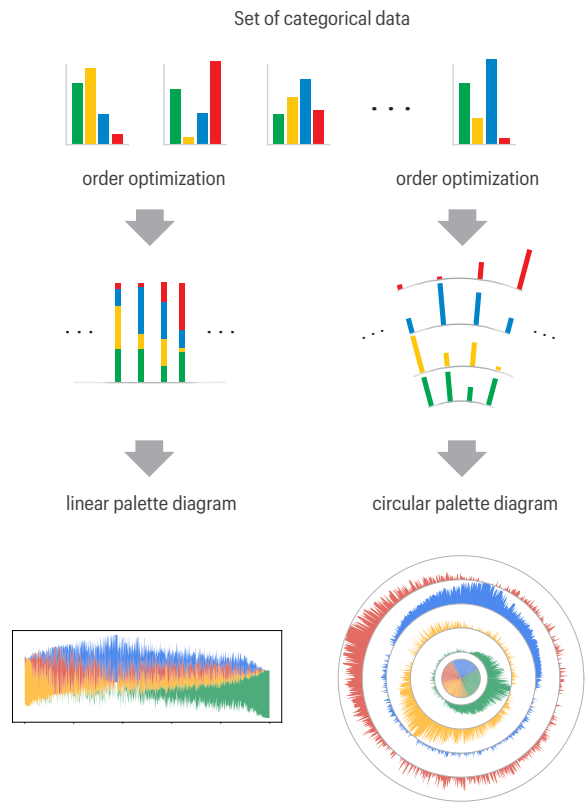}
  \caption{Schematic figures for the construction of palette diagrams  
  }
  \label{fig:Schematic}
\end{figure}

The package provides two types of palette diagrams (see Fig.~\ref{fig:Schematic} for illustrations): 

\begin{description}
\item[ Linear palette diagram ] This is a stream plot (or a stack plot with a varying axis), which is usually used for plotting time series data. 
Each categorical dataset is stacked vertically, and these stacked plots are aligned horizontally so that the neighboring datasets have similar vertical patterns. 
The concept of a linear palette diagram is described in further detail in \cite{NoguchiKawamoto2019}. 
\item[ Circular palette diagram ] 
A problem with the linear palette diagram is that it has boundaries. As the categorical dataset does not have an intrinsic order, considering a periodic boundary condition may be a better alternative. 
In a circular palette diagram, the order of the datasets is optimized on the polar coordinate, instead of the Cartesian coordinate. 
In addition, whereas the categorical data are stacked in the linear palette diagram, the data are presented in different layers in the circular palette diagram. 
The central part indicates the color of the dominant category within each categorical dataset (i.e., the maximum a posteriori estimate).
\end{description}

In both types of palette diagrams, the only required input is a set of categorical datasets. 

\section{Order optimization}
We denote the dimension of each categorical data as $K$ and the number of categorical data as $N$. 
The input for the palette diagram is expressed as $\{ \mbox{\boldmath $y$}_{i} \}_{i=1}^{N}$, where $\mbox{\boldmath $y$}_{i}$ ($1 \le i \le N$) is a column vector for the $i$th categorical data. 
$\{ \mbox{\boldmath $y$}_{i} \}$ is a set of $N$ data points in the $K$-dimensional space. 
We assume that every element in $\{ \mbox{\boldmath $y$}_{i} \}$ is nonnegative. 
Determining the order of the $N$ elements is a dimension reduction to a one-dimensional space. 

In the linear palette diagram, we use ISOMAP \cite{ISOMAP_Tenenbaum} for the order optimization, which is implemented in \textit{scikit-learn}. 
For the circular palette diagram, we use the following nonlinear embedding. 
We construct a $k$-nearest neighbor graph ($k = $\texttt{n\_neighbors}) from $\{ \mbox{\boldmath $y$}_{i} \}$ and generate an $N \times N$ distance matrix $D$ based on the shortest path length on the $k$-nearest neighbor graph. 
For the distance in the embedded space, we consider 
\begin{equation}\label{AngularDistance}
d_{ij} = 1 - \cos(\theta_{i} - \theta_{j}), 
\end{equation}
The angular coordinates $\{ \theta_{i} \}_{i=1}^{N}$ are optimized so that the following objective function is minimized: 
\begin{equation}\label{ObjectiveFunction}
L\left( \{ \theta_{i} \} \right) = \frac{1}{2} \sum_{i,j} \left( D_{ij} - d_{ij} \right)^{2}.
\end{equation}
The optimization is executed through the stochastic gradient descent method, that is, for a randomly selected pair of data elements $(i,j)$ $(i\ne j)$, $\theta_{i}$ and $\theta_{j}$ are updated as 
\begin{equation}\label{UpdateEqn1}
\theta_{i} \leftarrow \theta_{i} + \eta \sum_{j=1}^{N} \left( D_{ij} - d_{ij} \right) \sin\left(\theta_{i} - \theta_{j} \right).
\end{equation}
\begin{equation}\label{UpdateEqn2}
\theta_{j} \leftarrow \theta_{j} - \eta \sum_{i=1}^{N} \left( D_{ij} - d_{ij} \right) \sin\left(\theta_{i} - \theta_{j} \right). 
\end{equation}
where $\eta$ is the learning rate. 

In Sammon's nonlinear mapping (or the equivalent metric MDS) \cite{LeeVerleysen2007}, each term in the objective function is penalized by a weight $1/D_{ij}$. 
However, such a penalization is problematic because the case with $D_{ij}=0$ may not be rare in a set of categorical data. 
Moreover, empirically, Eq.~(\ref{ObjectiveFunction}) performs better than a variant of Sammon's nonlinear mapping in which the penalty is omitted for the pairs with $D_{ij}=0$. 
If we replace the angle-based distance (\ref{AngularDistance}) with the inner product $\theta_{i}\theta_{j}$ and convert the distance matrix $D$ to the Gram matrix after ``centering'' $\{ \mbox{\boldmath $y$}_{i} \}$, the above nonlinear embedding reduces to ISOMAP (pg.~76--77, \cite{LeeVerleysen2007}). 
As far as we have investigated, the implemented methods generally perform well, although we have also tried several other approaches for the order optimization, including neural networks \cite{NIPS2019_9015} and other manifold learning methods \cite{JMLRvandermaaten08a, Roweis2323}.

\section{Warning}
In the circular palette diagram, the values of each category are indicated in each layer. 
Notably, the more outer the layer (i.e., the larger the radius), the larger is the distance between neighboring elements on the diagram. 
Thus, the categorical data in the outer layers are amplified in terms of the area of the filled regions. 
Therefore, one should not (directly) compare the filled areas between different layers. 
Hence, we do not use the stack plot in the circular palette diagram. 
The order of layers is determined so that the category with a smaller $\sum_{i} y_{ik}$ is placed at the outer layer.

\begin{acknowledgments}
This study was supported by the New Energy and Industrial Technology Development Organization (NEDO), JST CREST Grant Number JPMJCR1912 (C.N.) and JSPS KAKENHI No. 18K18604 (T.K.).
\end{acknowledgments}

\bibliography{paper}

\end{document}